# Large-Area Electrodeposition of Few-Layer MoS$_2$ on Graphene for 2D Material Heterostructures

Yasir J. Noori[1]*, Shibin Thomas[2], Sami Ramadan[3], Danielle E. Smith[2], Vicki K. Greenacre[2], Nema Abdelazim[1], Yisong Han[4], Richard Beanland[4], Andrew L. Hector[2], Norbert Klein[3], Gillian Reid[2], Philip N. Bartlett[2] and C. H. (Kees) de Groot[1]*

[1]School of Electronics and Computer Science, University of Southampton, Southampton, SO17 1BJ, UK
[2]School of Chemistry, University of Southampton, Southampton, SO17 1BJ, UK
[3]Department of Materials, Imperial College, London, SW7 2AZ, UK
[4]Department of Physics, University of Warwick, Coventry, CV4 7AL, UK

**Abstract**

Heterostructures involving two-dimensional (2D) transition metal dichalcogenides and other materials such as graphene have a strong potential to be the fundamental building block of many electronic and opto-electronic applications. The integration and scalable fabrication of such heterostructures is of essence in unleashing the potential of these materials in new technologies. For the first time, we demonstrate the growth of few-layer MoS$_2$ films on graphene via non-aqueous electrodeposition. Through methods such as scanning and transmission electron microscopy, atomic force microscopy, Raman spectroscopy, energy and wavelength dispersive X-ray spectroscopies and X-ray photoelectron spectroscopy, we show that this deposition method can produce large-area MoS$_2$ films with high quality and uniformity over graphene. We reveal the potential of these heterostructures by measuring the photo-induced current through the film. These results pave the way towards developing the electrodeposition method for the large-scale growth of heterostructures consisting of varying 2D materials for many applications.

**Keywords:** Electrodeposition, non-aqueous, transition metal dichalcogenides, molybdenum disulfide, MoS$_2$, graphene, heterostructure, photodetector

## 1. Introduction

Transition metal dichalcogenide (TMDC) 2D materials such as molybdenum disulfide (MoS$_2$) are being developed to make ultrasensitive photodetectors, high on/off ratio transistors and sensitive sensors.[1–4] Mechanical measurements of MoS$_2$ have also shown that it is 30 times stronger than steel and can be deformed by up to 11%, making it one of the strongest semiconductor materials for applications in flexible electronics.[5] Heterostructures based on 2D materials can enable new technologies due to the unique properties of their atomic layer constituents.[6–8] Specifically, integrated photodetector heterostructures based on stacking TMDC on graphene have been shown to have high speed, responsivity and gain.[9–12] These devices commonly use graphene for their electrical contacts, due to its exceptional electrical conductivity and optical transparency for (opto-) electronic technologies.

Wafer scale production of graphene has been enabled through chemical vapour deposition (CVD).[13,14] However, a major obstacle that hinders the development of MoS$_2$ based heterostructures is its low scalability. Several methods have been developed to produce MoS$_2$ such as mechanical, liquid and chemical exfoliation,[15–18] chemical vapour deposition,[19,20] plasma sputtering,[21] solution based thermal decomposition and pulsed laser deposition.[22,23] However, each of these methods has certain limitations that prevent it from producing large, uniform and continuous monolayers with a high yield, low cost and controllable atomic layer thickness. For example, spatial control of material production using mechanical exfoliation is very difficult and industrially unscalable. The CVD method produces triangularly shaped crystals that randomly locate and seed on the substrate making the production of continuous layers a difficult task. The CVD and sputtering methods require vacuum equipment and dedicated infrastructure which makes them relatively expensive. An alternative method that has rarely been investigated is electrodeposition. Atomic layers are deposited through this bottom up method by applying an electric potential to an electrode in a solution containing the precursor of the material to be deposited. Site selective material growth in the nanometre scale can be achieved by patterning the electrodes and/or coating the electrodes with a patterned non-conductive layer such as SiO$_2$.[24] In comparison to the aforementioned methods for depositing MoS$_2$, electrodeposition is potentially capable of producing wafer-scale continuous films of 2D materials for reasonably low costs. In addition, a unique advantage of electrodeposition is that it is not a line-of-sight deposition method and can be utilised to conformally coat complex three dimensional structures including patterned structures of high aspect ratios. This method also offers the prospect for integrating 3D shaped graphene nanosheets with other 2D materials to make van der Waal hybrid materials for superior performance and promising applications in various energy fields.[25] Techniques such as plasma sputtering, CVD and atomic layer deposition (ALD) are normally carried out in harsh environments or at very high temperatures that destroys existing 2D materials on the substrate such as graphene. On the other hand, electrodeposition does not affect its quality as it can be performed at room temperature and pressure.[26–31] To the best of our knowledge, the CVD and ALD methods have never demonstrated large scale and continuous film growth of MOS$_2$ over graphene. Limitations of electrodeposition, however, are low crystallinity deposits, the deposition on only conductive materials and the possible incorporation of impurities from the electrolyte.

A few works have previously attempted to electrodeposit bulk MoS$_2$ using an aqueous solution over glassy carbon, polytetrafluoroethylene and conductive glass.[32–36] However, building 2D heterostructures for device applications requires that the deposition be done over other 2D materials such as graphene, a common electrical bottom contact material for device structures. One report has previously shown the simultaneous deposition of thick MoS$_3$ and MoS$_2$ films on graphene via galvanostatic two-electrode aqueous deposition.[37] The films were then converted to MoS$_2$ by an annealing step. However, the main disadvantage of the galvanostatic deposition method is that

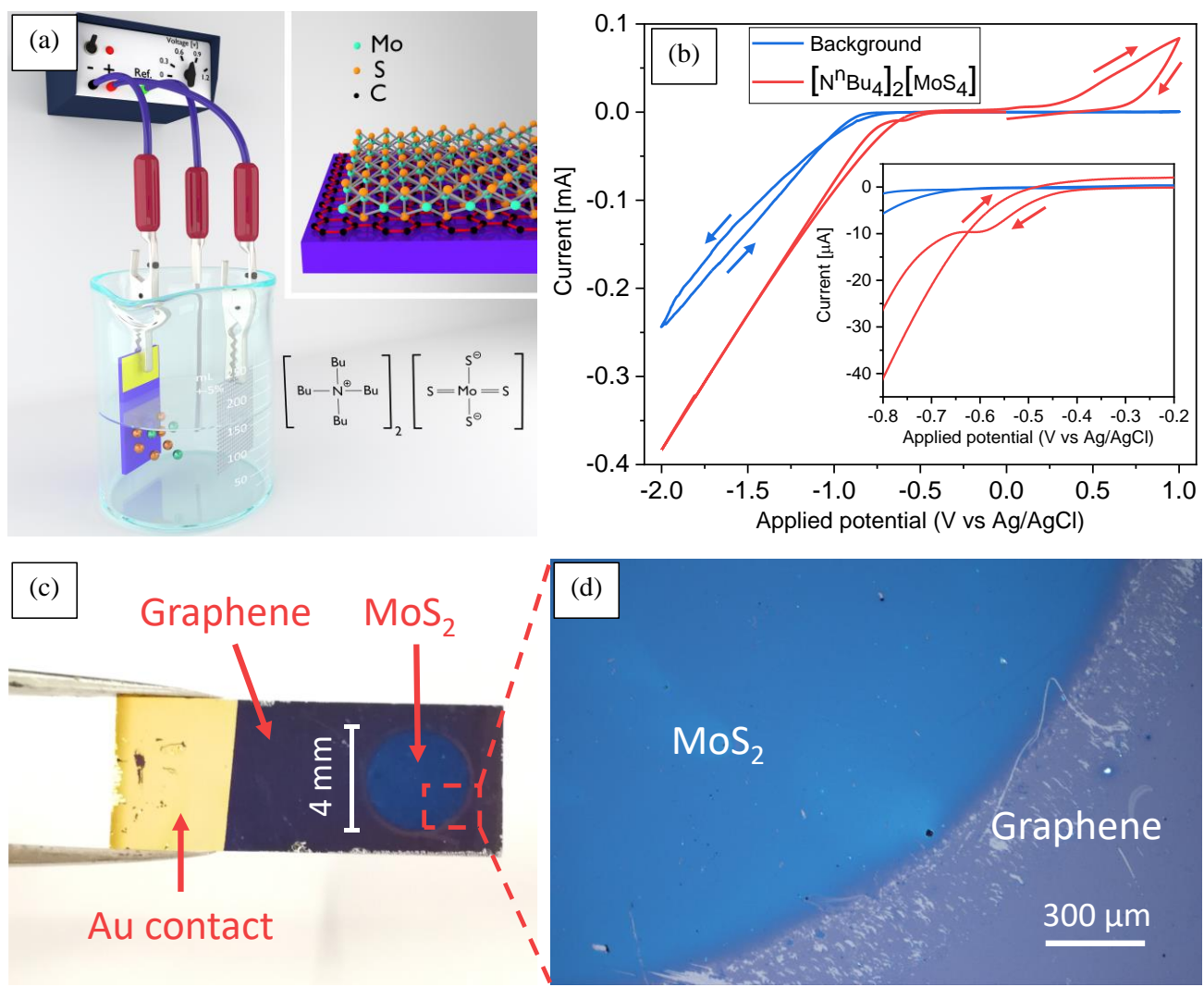

Figure 1 (a) A schematic illustration of a three-electrode electrodeposition setup with a substrate, a Pt gauze and a Ag/AgCl wire as the working, counter and reference electrodes, respectively. The schematic also shows an illustration of a $MoS_2$/graphene heterostructure and the precursor used in this work. (b) A CV scan comparison between a solution containing a 5 mM $[N^nBu_4]_2[MoS_4]$ and 0.2 M $(CH_3)_3NHCl$ with the supporting electrolyte $[N^nBu_4]Cl$ in $CH_2Cl_2$ (red curve) and the background solution which only contains the supporting electrolyte in $CH_2Cl_2$. (c) A real photograph of the Si/SiO$_2$ substrate after electrodeposition, showing the area where the $MoS_2$ was grown over graphene and a microscope image (d) of the $MoS_2$ and graphene regions.

it does not allow controllability over the electrodes' potentials, which becomes essential in controlling the deposition of atomically thick materials. In addition, the electrochemical potential window is limited in aqueous solution to approximately 1.3 V by electrolysis of water. This limits the uses of the aqueous electrodeposition method to depositing only materials that reduce at moderate potentials. $MoS_2$ is an electrocatalyst for hydrogen evolution, hence the deposition of $MoS_2$ facilitates the evolution of $H_2$ which makes the solution at the electrode basic, further complicating the process. Electrodeposition in non-aqueous solvents using a three-electrode electrodeposition setup is an alternative to overcome these limitations.[24,38–40] A commercially available precursor for electrodeposition is $[NH_4]_2[MoS_4]$. However, this precursor is not compatible with non-aqueous solvents such as dichloromethane ($CH_2Cl_2$) or acetonitrile ($CH_3CN$).

In this work, we show for the first time that the electrodeposition method can produce few-layer thin $MoS_2$ films using a graphene electrode. To the best of our knowledge, this work presents the thinnest and most uniform $MoS_2$ film ever produced via the electrodeposition method.[32,37,41] We are demonstrating the electrodeposition of $MoS_2$ over graphene using tetrabutylammonium tetrathiomolybdate $[N^nBu_4]_2[MoS_4]$ as a single source precursor that is compatible with $CH_2Cl_2$ and makes a better replacement to the more common $[NH_4]_2[MoS_4]$ precursor for non-aqueous electrodeposition. Using a variety of characterisation techniques, we have shown that our $MoS_2$ is of high quality and semiconductor, which has allowed us to do photo absorption experiments based on the $MoS_2$/graphene heterostructure.

## 2. Experimental Section

Graphene monolayers were grown by CVD on a copper foil. Details of the growth, a photograph of graphene coated substrate, electrical and atomic force microscopy (AFM) measurements can be found in supporting information figure S1. The graphene was spin coated with poly methyl methacrylate (PMMA) and wet transferred onto the target Si substrate by wet etching the copper foil and slowly scooping the floating PMMA/graphene film by the substrate. The PMMA layer was later dissolved in acetone and rinsed off with isopropyl alcohol (IPA). To improve the electrical contact a Cr (10 nm)/Au (190 nm) layer was thermally evaporated on part of the graphene film away from the electrodeposition area. $[N^nBu_4]_2[MoS_4]$ was synthesised in-

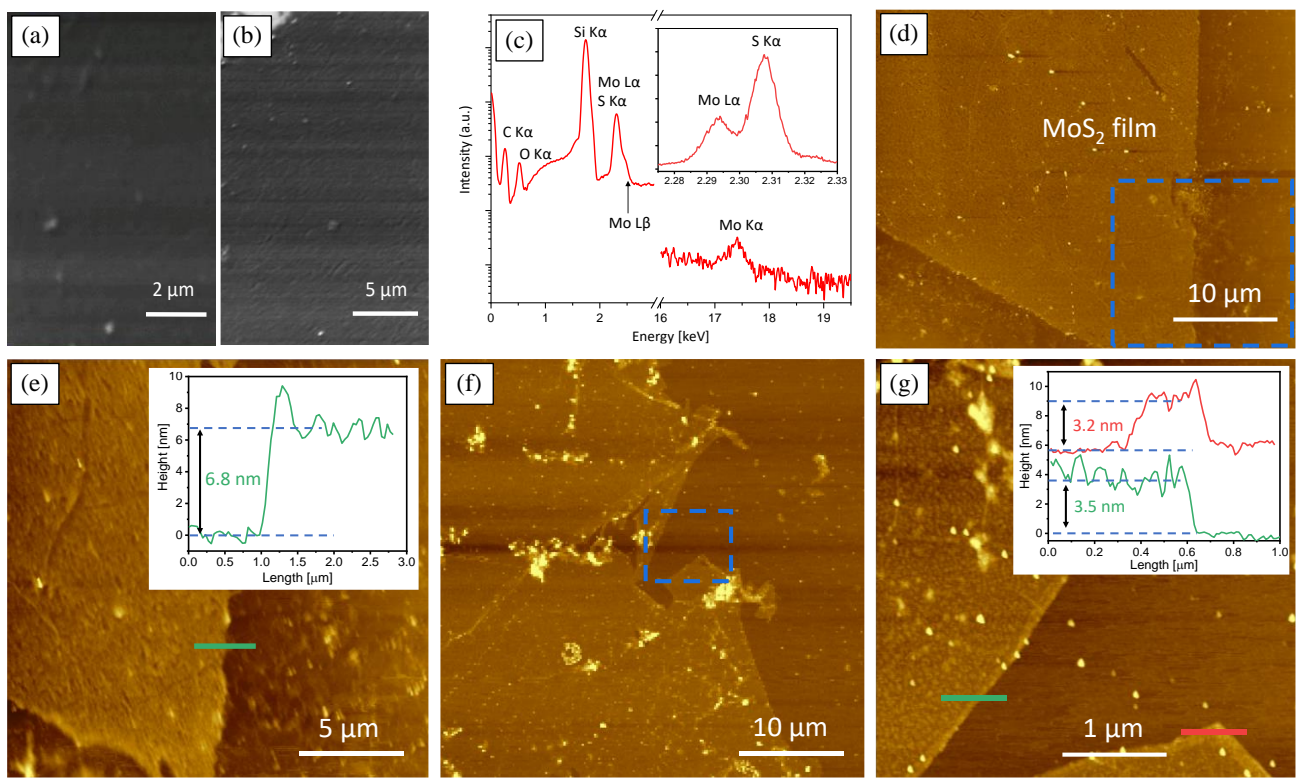

Figure 2 (a) and (b) Scanning electron micrographs of the uniform electrodeposited film of MoS$_2$ before and after annealing (c) Energy dispersive and wavelength dispersive (figure inset) X-ray spectroscopy graphs showing the presence of Mo and S in the film with 1:2 ratio. AFM images of two films deposited at different times for 90 s (d and e) and 10 s (f and g) with insets showing the thickness of the MoS$_2$/graphene film measured relative to the substrate. The blue dashed squares represent the magnified AFM scans presented in (e and g) for the two films, respectively.

house to function as a single source precursor that can deliver both Mo and S, and has excellent solubility in CH$_2$Cl$_2$ compared to [NH$_4$]$_2$[MoS$_4$].[42] [N$^n$Bu$_4$]Cl was used as the supporting electrolyte and trimethylammonium chloride (CH$_3$)$_3$NHCl was used as the proton source. The electrodeposition process, including the electrolyte preparation was carried out inside a glovebox equipped with a nitrogen recirculating system to maintain the O$_2$ and H$_2$O levels below 10 ppm. The deposition was performed inside a three-electrode electrochemical cell that employs a Pt gauze as a counter electrode and a Ag/AgCl (0.1 M [N$^n$Bu$_4$]Cl in CH$_2$Cl$_2$) reference electrode as shown in Figure 1 (a). The CH$_2$Cl$_2$ used in this work was previously dried and degassed by refluxing with CaH$_2$ before distillation to minimise its water content. The moisture content was measured by Karl-Fischer titration to be ca. 18 ppm.

Cyclic voltammetry (CV) scans were carried out to study the electrochemical behaviour of the electrolyte with the graphene working electrode. The electrolyte and background CV scans were performed on two different graphene coated substrates on separate days, using a newly prepared electrolyte each time, see Figure 1 (b). A scan is sweeping the voltage from 0 to -2.0 V, -2.0 to 1.0 V and 1.0 to 0 V. The indicative arrows on the observed current curves show the direction of the sweep. The CV scans show a clear dip in the current at around -0.6 V which can be clearly observed in the magnification inset of the figure. This dip was confirmed to be caused by the electroreduction of [MoS$_4$]$^{2-}$ ions to MoS$_2$. We confirmed the deposition of material at this potential by polarising the graphene electrode at -0.8 V for 30 minutes which resulted in a thin film uniformly covering the electrode. The CV scan on graphene is compared to that on TiN and HOPG in the supporting information Figure S2. The overall resistive behaviour of the CV scan is due to the iR drop across the electrolyte.

The electrodeposition of MoS$_2$ is achieved by fixing the potential applied to the working electrode at -0.8 V and varying the deposition time to control the thickness of the deposited material.[41] The reaction formulae at the working and counter electrode are respectively as follows

$$\text{MoS}_4^{2-} + 2e^- + 4H^+ \rightarrow \text{MoS}_2 + 2H_2S \quad (1)$$

$$\text{MoS}_4^{2-} \rightarrow \text{MoS}_3 + \tfrac{1}{8}S_8 + 2e^- \quad (2)$$

During the electrodeposition process, the substrate is placed inside a sealed container to ensure that MoS$_2$ is deposited within a predefined area (4mm circle) to prevent materials from depositing at the sides and back of the substrate. We expect this approach to be easily scalable to large wafer scales with a larger reactor setup and large-area graphene. After the deposition, the sample is rinsed with pure CH$_2$Cl$_2$ and left to dry inside the glove box. Figure 1 (c) shows a photograph of an as-deposited large-area MoS$_2$ films with a thickness of 5.8 nm following a 90 s deposition, demonstrating a large colour contrast to pristine graphene. Figure 1 (d) depicts a microscope image of the deposited MoS$_2$ film, showing the uniform colour of the MoS$_2$ film to give an indication of the uniformity and continuity of the deposited material at the millimetre scale. The damage on the graphene at the border with the MoS$_2$ film is caused by the reactor's O-ring which seals other areas of the substrate from being exposed to the electrolyte solution. This problem can be easily solved for device applications by patterning a protective insulating layer such as a photoresist.

The as-deposited MoS$_2$ is amorphous, hence an annealing step was introduced in the process to crystallise the film. The sample annealing was done inside a tube furnace initially at 100 °C for 10 minutes followed by a 500 °C for two hours at 0.1 mbar. The

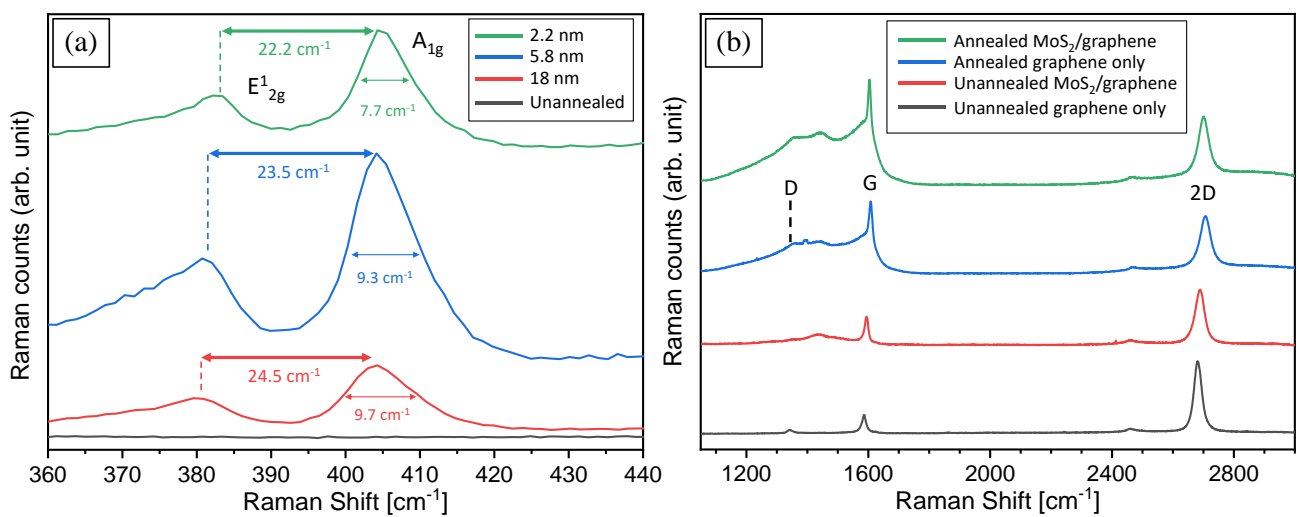

Figure 3 (a) Raman spectra for electrodeposited MoS$_2$ showing that the separation between the A$_{1g}$ and E$^1_{2g}$ peaks reduces as the deposition thickness is reduced. An unannealed sample shows no signal beyond the instrument's noise level as depicted by the black curve. (b) Raman spectra of graphene showing that the 2D peaks remains intense after annealing with an increase in the G peak's intensity and the appearance of peaks near the defects window at around 1350 cm$^{-1}$. The curves are on the same scale but separated for clarity.

annealing took place in a sulfur rich environment produced by placing 0.1 g of sulfur powder source inside the furnace.

## 3. Results and Discussions

**Energy/Wavelength Dispersive X-ray Spectroscopy**

Scanning electron microscopy (SEM) images of the deposited film before and after annealing are shown in Figure 2 (a and b). Energy dispersive X-ray spectroscopy (EDX) and wavelength dispersive X-ray spectroscopy (WDX) were used to confirm the deposition of MoS$_2$ on the graphene and to measure its atomic ratios. The EDX shows the presence of C, O, Si, Mo and S, that are expected to come from graphene, PMMA residue, the SiO$_2$ substrate and the deposit. This indicates that the electrodeposition is a clean method that can deposit high purity 2D materials. Other elements that are present in the electrolyte such as chlorine were not found in the deposit. Due to the spectral overlap between the Mo-L$\alpha$ (2.293 keV) and the S-K$\alpha$ (2.307 keV) X-ray emission lines, WDX was used to resolve the Mo and S spectra and calculate the film's material composition, as shown in the inset of Figure 2 (c). The S/Mo ratio was measured to be 2 ± 0.1. In both measurements, a commercial MoS$_2$ crystal was used as a calibration standard to calibrate the tools. The measurements were repeated after annealing and a negligible change was found in the S/Mo ratio.

**Atomic Force Microscopy**

Atomic force microscopy was used to measure the thickness of the electrodeposited MoS$_2$ after annealing. Figure 2 (d) and its magnified view (e) shows the height profile, relative to the substrate, of a film deposited for 90 s. Figure 2 (f) and its magnified view (g) show the height profiles of a film deposited for 10 s. The AFM height profiles for the two attempts show total thicknesses of ~6.8 nm and ~3.2 nm, respectively. These thicknesses correspond to the thickness of the electrodeposited MoS$_2$ film plus the thickness of the graphene electrode. AFM measurements of bare graphene shows that our graphene has a thickness of ~1.0 nm as shown in the supporting information Figure S1. Therefore, the electrodeposited MoS$_2$ films have thicknesses of ~5.8 and ~2.2 nm, respectively. Hence, the 2.2 nm film is expected to contain a bilayer or trilayer MoS$_2$ film. The small particles at the top of the film shown in Figure 2 (g) may also indicate that there is a "uniform" film and a partially grown layer that is stopped at an intermediate seeding state. The supporting information Figure S3 shows magnified AFM images of that film and another deposited for 10 minutes with a thickness of 18 nm. AFM characterisation of these films has shown that they have significantly better film continuity and uniformity than any other electrodeposited MoS$_2$ films from previous works.[32–36,43]

**Raman Spectroscopy**

Raman spectroscopy is a common method used to characterise TMDC materials. In this work, the presence of MoS$_2$ on the substrate, its thickness and degree of crystallinity were investigated by measuring the Raman scattering of a 532 nm laser at room temperature. Light excitation and collection were implemented using a 50x objective to reduce the measurement area to ~1 µm$^2$. Raman measurements of MoS$_2$ primarily report the study of the E$^1_{2g}$ and A$_{1g}$ scattering peaks. It has been established by several studies that the separation between the two peaks is strongly dependent on the number of stacked MoS$_2$ layers, up to "bulk" films.[44–48] The separation was found to increase as the number of layers is increased, and vice versa. The blue shift in the A$_{1g}$ peak with increasing the number of layers is commonly attributed to higher atomic vibration force constants caused by interlayer Van der Waals forces. The relatively smaller red shift in the E$^1_{2g}$ peaks, on the other hand, is attributed to stacking-induced structural changes or long-range Coulombic interlayer interactions in multilayer MoS$_2$. In this work, several electrodeposited films were studied with Raman spectroscopy. Figure 3 (a) shows the Raman spectra obtained from three annealed films with different thicknesses and a fourth spectrum from an unannealed film. The electrodeposited MoS$_2$ films did not show any Raman signature before annealing, indicating that the as-deposited material is amorphous and that the annealing step is required to crystallise the deposits. The separation between the two Raman peaks for MoS$_2$ were found to reduce when the film thickness determined by AFM is reduced as expected, matching previous reports in literature. The 18 nm thick film showed a peak separation of 24.5 cm$^{-1}$, this separation is conventionally abbreviated by "bulk" MoS$_2$ indicating that the material is more than ten monolayers thick. The thinner film characterised in Figure 2 (d) showed a peak separation of 23.5

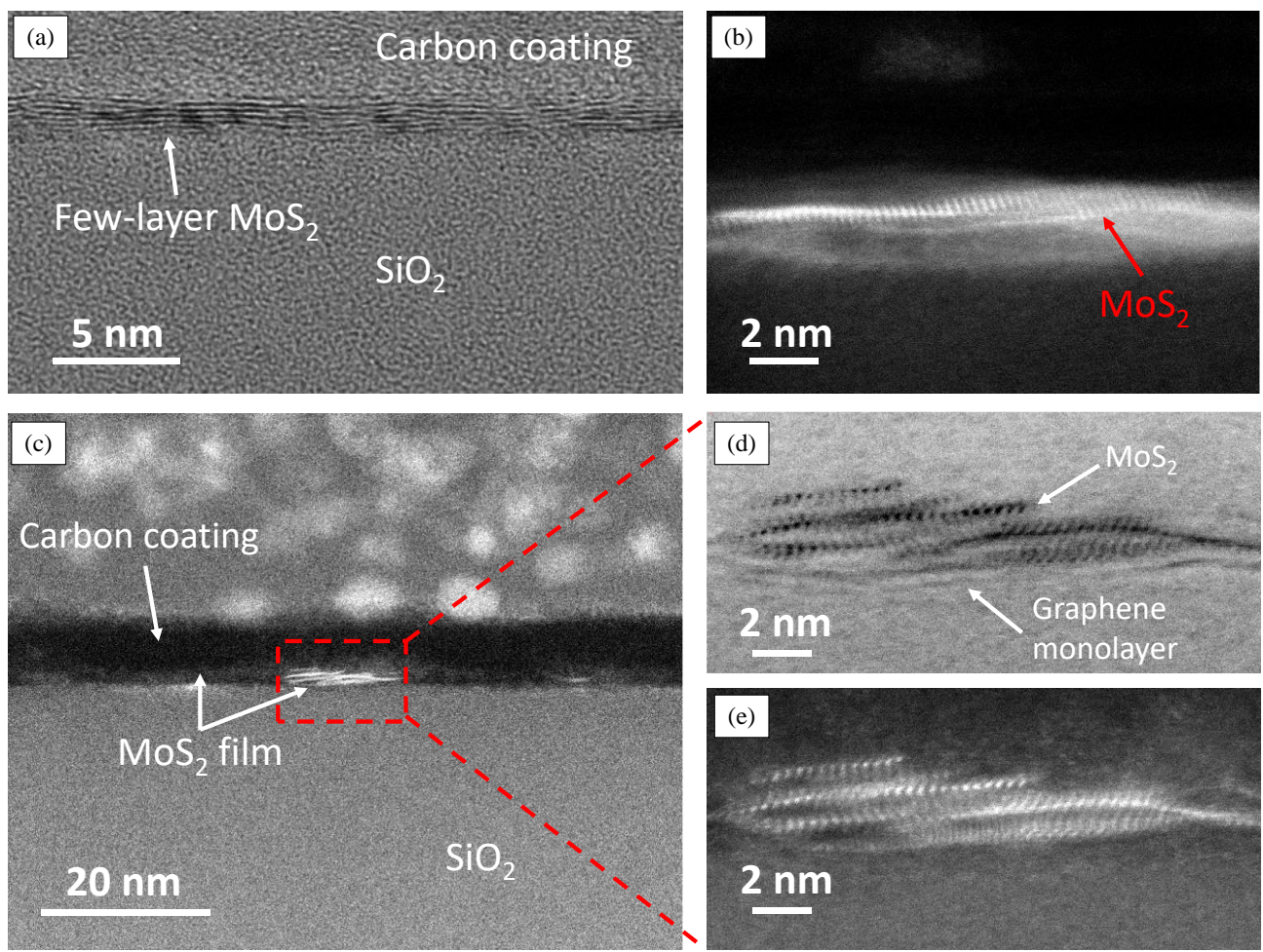

Figure 4 (a) Bright Field (BF) TEM and (b) Annular Dark Field (ADF) TEM images of an electrodeposited $MoS_2$ film showing few stacked layers of the material and a periodic pattern associated to the crystal. (c) ADF-TEM image of a second electrodeposited $MOS_2$ film and its corresponding magnified BF (d) and ADF (e) TEM images showing the periodic crystal pattern and a graphene monolayer clearly visible underneath.

$cm^{-1}$, indicating a grown film of "few-layer" $MoS_2$. A series of scans across a large film area can be found in the supporting information Figure S4, showing a strong overlap between the peaks. The thinnest film, characterised in Figure 2 (f) shows a peak separation of 22.2 $cm^{-1}$, indicating a bilayer $MoS_2$ film[44,46]. The full width half maxima (FWHM) of the $A_{1g}$ and $E^1_{2g}$ peaks were also found to decrease with decreasing the number of layers, which may indicate a higher level of crystallinity. Figure 3 (b) shows four Raman spectra of bare and $MoS_2$ coated graphene, before and after annealing. The intensity ratio of the 2D and G peaks, $I_{2D}/I_G$, following the electrodeposition process was found to be ~3.7. This means that the electrodeposition process has little effect on the crystallinity of graphene. A very small peak around the defect related D band around 1350 $cm^{-1}$ was found with a small increase in the Raman shift for the 2D and G peaks, which may be attributed to graphene doping.[49] However, the $I_{2D}/I_G$ ratio was reduced to ~1.0 and ~0.9 after annealing for graphene only and graphene/$MoS_2$ areas, respectively. The hump near the defect band, and the further increase in the Raman shift for the 2D and G peaks is most likely due to sulfur induced hole doping and compressive stress in graphene.[50] It is expected that lowering the annealing temperature, time and sulfur level in the chamber will reduce graphene doping, however, this is also expected to produce lower crystallinity $MoS_2$. Therefore, further optimisation of the annealing temperature may be needed to minimise the sulfur doping of graphene that results from this process.[51] It is worth emphasising that the quality of graphene reported here, which had undergone an electrodeposition process, remains significantly better than those that had undergone a plasma sputtering process from previous reports.[26–30]

**Transmission Electron Microscopy**

Cross-sectional TEM imaging of the film following a lamella process and a protective carbon coating was performed using Bright field (BF) and Annual Dark Field (ADF) modes. The ADF mode allows distinguishing $MoS_2$ from its surrounding $SiO_2$ and C due to the atomically heavy Mo, which causes greater electron scattering, thus observed with higher brightness. Ordered stacking of few-layer electrodeposited $MoS_2$ was observed in Figure 4 (a). A periodic pattern corresponding to the atomic arrangement of the material was also visible in Figure 4 (b). In another sample, a $MoS_2$ film was observed as a faint bright layer between the protective carbon and the $SiO_2$ as shown in Figure 4 (c). An area of the film was focused on and atomic layers of $MoS_2$ on top of the graphene electrode were clearly visible as shown in Figure 4 (d and e).

**X-ray Photoelectron Spectroscopy**

X-ray photoelectron spectroscopy (XPS) measurements were performed before and after annealing to study the chemical environments and composition of the Mo and S atoms in the films. Figure 5 shows the Mo 3d (a and c) and S 2p (b and d) spectra for the exact same area before and after annealing. The

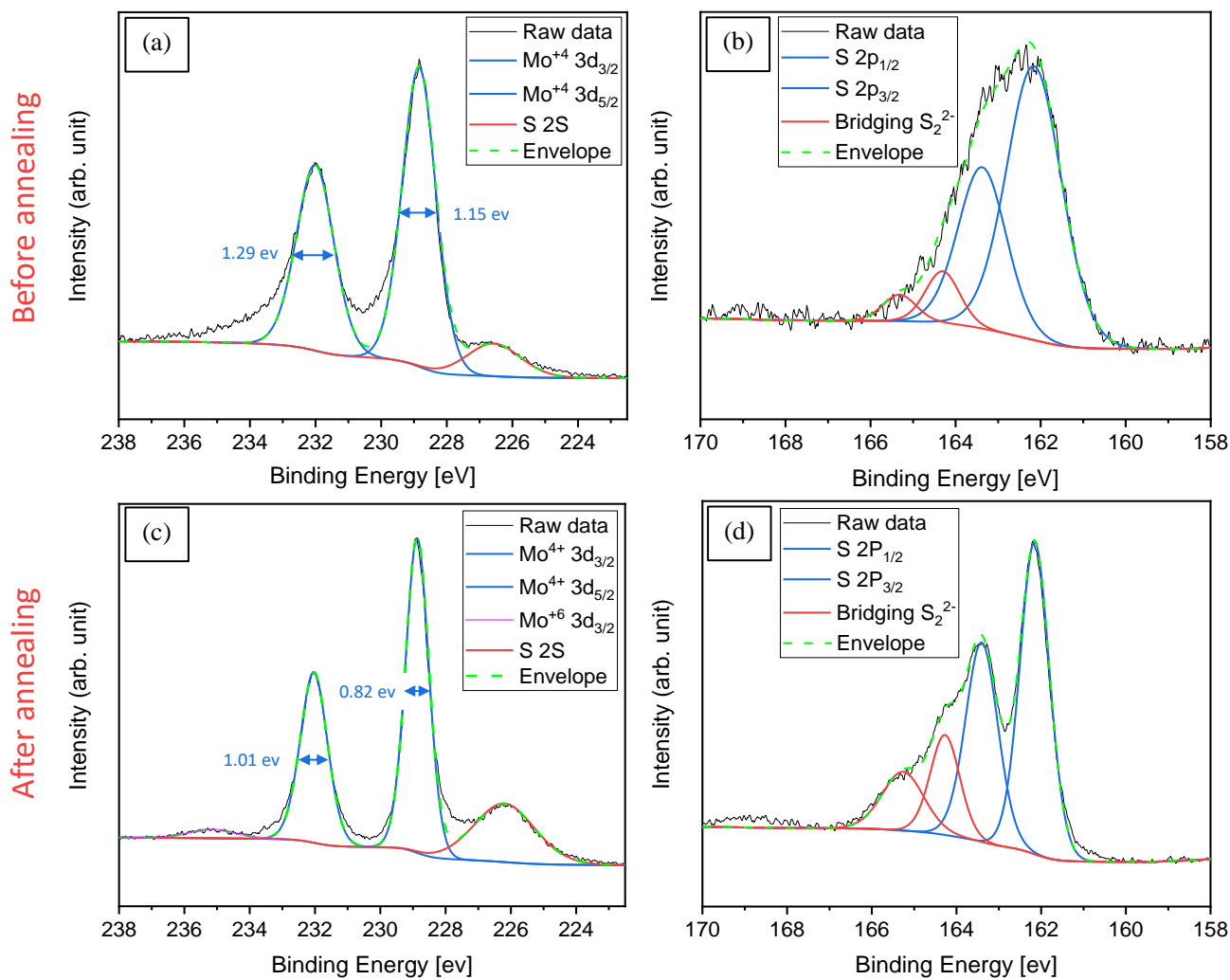

Figure 5. XPS spectroscopy measurements of the Mo3d (a) and S 2p (b) energy ranges before and after (c and d) annealing. The graphs show the clear improvement in the material crystallinity after annealing as abbreviated by the sharper and more intense emission peaks. The y-scale is 1.5 times higher after annealing.

Mo 3d spectra can be resolved primarily into a doublet at 228.9 and 232.0 eV, corresponding to the $Mo^{+4}$ $3d_{3/2}$ and $3d_{5/2}$, respectively. The S 2s peak can also be found at 226.2 eV in the same spectra. No evidence of $Mo^{+6}$ was detected before annealing. After annealing, The $Mo^{+4}$ 3d peaks were found to become sharper and more intense. The reduction in the full width half maximum after annealing is an indication of increased crystal order in the material.[52] Figure 5 (c) also shows a very small peak at 235.3 eV which may be related to a $Mo^{+6}$ $3d_{3/2}$ oxidation peak. The two S 2p emission peaks at 161.7 and 162.9 eV corresponding to S $2p_{3/2}$ and $2p_{1/2}$, respectively, evolved to become much more intense and prominent after annealing as shown in Figure 5 (b and d). This was also accompanied by a great reduction in the FWHM, which may also be attributed to the increased crystal order of the material. In addition, two extra peaks were found at 163.8 and 164.8 eV that are believed to correspond to bridging $S_2^{2-}$ and/or apical $S^{2-}$ ligands, as shown in Figure 5 (d). In addition, a very small peak, probably corresponding to $[SO_4]^{2-}$, was observed at around 169 eV. All the XPS results are consistent with the previously reported measurements in the literature.[36,52–54] The S/Mo composition ratio of the film was also quantified via XPS to be 2.0 ± 0.1 following annealing and a short ion etch of the top few atomic layers. A wide energy range XPS scan can be found in supporting information Figure S5. The wide scan shows significantly lower oxygen content in the film compared to that in the literature[43] which is expected to arise due to the presence of oxygen in aqueous electrolytes.

## 2D Material Heterostructure photoresponsivity

We tested the photoresponse of our large-area electrodeposited $MoS_2$ film through graphene at room temperature and ambient air using a 532 nm laser source with ~ 1 mW power, see Figure 6 (a). Two electrical probes were placed on the $MoS_2$ layer and the Au back contact to act as the positive and negative terminals of the device, respectively. In a photodetector device, the electrical contact through the $MoS_2$ film is typically symmetric. However, for the purpose of testing the film's semiconductor optical absorption and induced current, the positive probe directly contacted a 25 nm thick $MoS_2$/graphene layer without the use of a graphene/Au contact pad. Because the latter may short the electrical path to the underlying graphene layer through cracks or voids in the $MoS_2$ film. A voltage sweep between -1 and 1 V was applied to the device to characterise its electrical response in the dark and under illumination as shown in Figure 6 (b). The IV characteristics of the device have shown a Schottky behaviour with up to an order of magnitude change in current between -1 and 1 V, which presumably arises at the $MoS_2$/graphene junction. Under illumination, the photocurrent (red line) was found to increase by three-fold compared to that for the dark current (blue line) at positive voltage biases. On the other hand, the photocurrent was found to be only marginally larger than the dark current at negative voltage biases. Figure 6

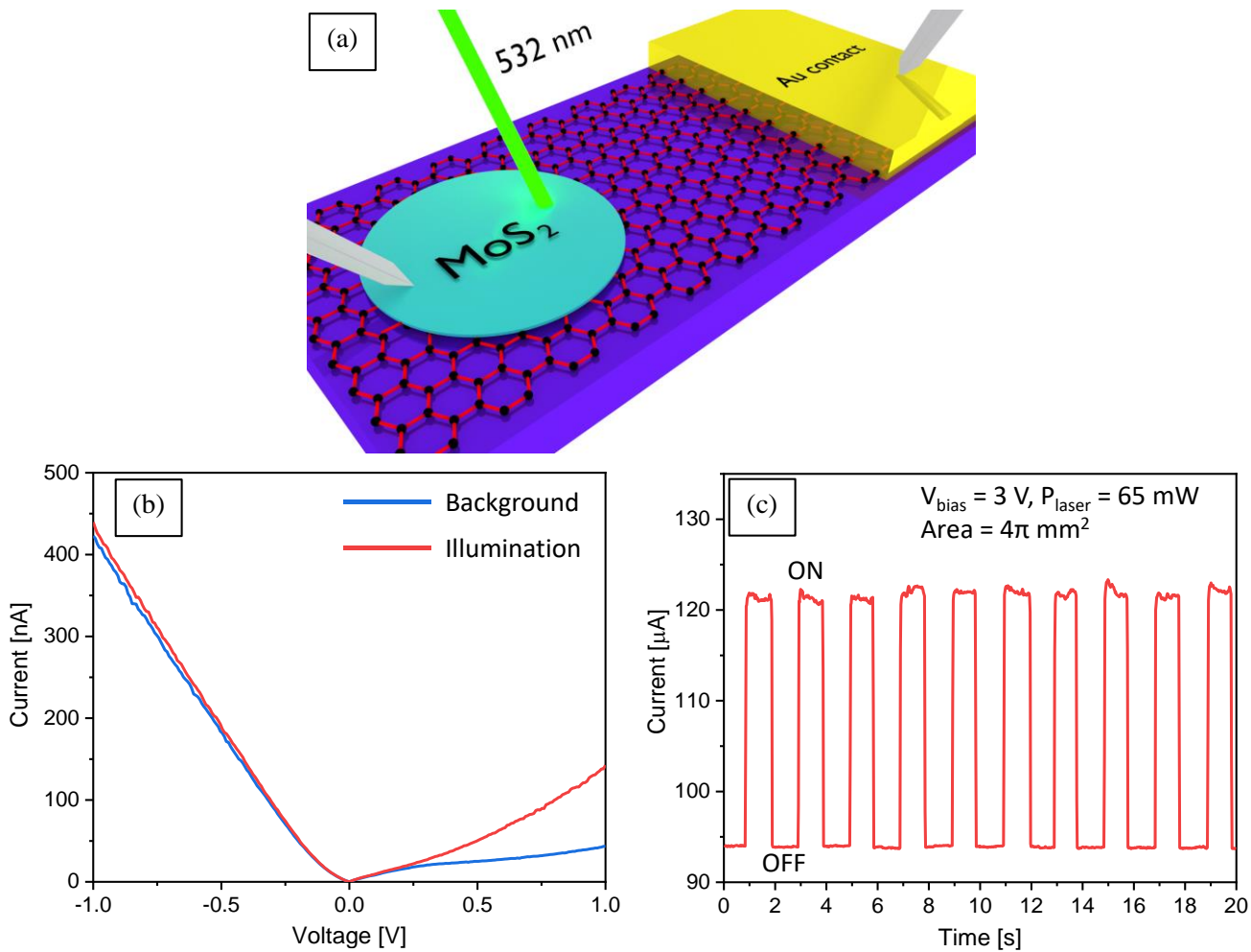

Figure 6 (a) A schematic illustration of the electrodeposited MoS$_2$/graphene heterostructure photodetector. The device has a MoS$_2$ film thickness of ~25 nm contacted via graphene using a Cr/Au electrode of 10/190 nm thickness and excited with a 532 nm laser with P ~1 mW. (b) IV characterisation of the device showing a clear current increase with illumination (red line) compared to that without illumination (black line). (c) photo-illumination cycles of another sample showing the switching of induced photocurrent with the switching of a laser source.

(c) presents the photocurrent induced in a sample via a pulsing 65 mW laser using a voltage bias of 3 V. Similar measurements performed on both, pristine graphene areas and unannealed MoS$_2$ films showed no photoresponsive behaviour.

The photoresponsivity of the film can be calculated using the following equation:

$$Photoresponsivity = \frac{I_{ph}}{P_{laser}}$$

Where $I_{ph}$ is the photoinduced current. $I_{ph}$ can be found by subtracting the device current under illumination from that in the dark ($I_{ph} = I_{light} - I_{dark}$), and P is the incident laser power. In this work, the maximum photoresponsivity at $V = 1\ V$ was found to be approximately $0.5\ mA\ W^{-1}$. The preliminary responsivity obtained from this film is much smaller than previously reported but confirms that the material is photosensitive.[9,11] This low responsivity is due to first the large area separation between the top and bottom area contacts. Second due to the large film area and millimetre size (unfocused) laser spot size that was applied to the film without the use of a focusing objective lens. Our future work involves fabricating practical micron size MoS$_2$/graphene photodetector devices with symmetric contacts and using an objective lens to focus the laser spot to a micron spot size for testing.

## 4. Conclusions

We reported the successful growth of large-scale and continuous MoS$_2$ film on graphene via non-aqueous electrodeposition. We used an in-house synthesised precursor that is designed to be compatible with dichloromethane CH$_2$Cl$_2$ for non-aqueous electrodeposition. The electrochemical behaviour of the precursor was investigated via cyclic voltammetry studies and a reduction potential around -0.6 V was found to be suitable for the electrodeposition of MoS$_2$. The material's purity and compositional ratio were measured via energy and wavelength dispersive X-ray spectroscopies, and the S/Mo ratio was found to be 2.0 ± 0.1. AFM, Raman, TEM and XPS measurements were used to measure the thickness of the MoS$_2$ film and characterise its evolution after annealing. We found that an annealing step is required to improve the crystallinity of the film following electrodeposition. Raman spectroscopy confirmed that the electrodeposition method can preserve the quality of the graphene electrode for future TMDC/graphene heterostructures in contrast to other methods such as plasma sputtering. The MoS$_2$/graphene heterostructure was subsequently tested for photodetection application and the electrodeposited film has shown a clear photoresponse indicating the material is in the semiconductor phase and the method's potential for future (opto-) electronic applications.

Electrodeposition is a cost-effective, simple, and flexible method that has great potential to be scaled to wafer sizes to enable the deposition of several layers of different 2D materials for heterostructure devices. This is in contrast with more commonly used methods such as CVD and mechanical exfoliation. Other materials that are likely to be feasible to grow via electrodeposition on graphene include $WS_2$, $MoSe_2$, $WSe_2$, $MoTe_2$, $WTe_2$, etc. The non-aqueous electrolyte also allows a larger electrochemical potential window providing a more flexible foundation to future incorporation of dopants, and heterostructure growth of different 2D semiconductors. This paves the way towards making the electrodeposition method a key method in the manufacturing of 2D material heterostructure devices for applications in flexible light emitters, photodetectors, transistors, sensors and batteries as examples.

## Author contributions

YJN proposed the idea and coordinated the project; ST and PNB worked on the electrodeposition experiments; SR and NK worked on the graphene growth and transfer; DES, VG, ALH and GR worked on the precursor preparation; YH and RB performed the TEM imaging; YJN, NA and CHDG worked on the material characterisation. YJN wrote the manuscript with contributions from all authors.

## Notes

The authors declare no competing interest.

## Author information

**Corresponding Authors**

**Yasir J. Noori** – School of Electronics and Computer Science, University of Southampton, Southampton, SO17 1BJ, UK
Email: y.j.noori@southampton.ac.uk ; yasir.noori@gmail.com

**C. H. (Kees) de Groot** – School of Electronics and Computer Science, University of Southampton, Southampton, SO17 1BJ, UK

Email: chdg@southampton.ac.uk

**Other Authors**

**Shibin Thomas** – School of Chemistry, University of Southampton, Southampton, SO17 1BJ, UK

**Sami Ramadan** – Department of Materials, Imperial College, London, SW7 2AZ, UK

**Danielle E. Smith** – School of Chemistry, University of Southampton, Southampton, SO17 1BJ, UK

**Vicki K. Greenacre** – School of Chemistry, University of Southampton, Southampton, SO17 1BJ, UK

**Nema Abdelazim** – School of Electronics and Computer Science, University of Southampton, Southampton, SO17 1BJ, UK

**Yisong Han** – Department of Physics, University of Warwick, Coventry, CV4 7AL, UK

**Richard Beanland** – Department of Physics, University of Warwick, Coventry, CV4 7AL, UK

**Andrew L. Hector** – School of Chemistry, University of Southampton, Southampton, SO17 1BJ, UK

**Norbert Klein** – Department of Materials, Imperial College, London, SW7 2AZ, UK

**Gill Reid** – School of Chemistry, University of Southampton, Southampton, SO17 1BJ, UK

**Phillip N. Bartlett** – School of Chemistry, University of Southampton, Southampton, SO17 1BJ, UK

## Acknowledgements

The research work reported in this article was financially supported by the Engineering and Physical Sciences Research Council (EPSRC) through the research grant EP/P025137/1 (2D layered transition metal dichalcogenide semiconductors via non-aqueous electrodeposition) and the programme grant EP/N035437/1 (ADEPT - Advanced devices by electroplating). We also thank Kath Leblanc and Nikolay Zhelev for their technical assistance in this work and Omar Abbas for the useful discussions.

## Associated Contents

**Supporting Information**

Experimental details for the graphene electrodes along with the electrochemical behaviours of the electryolte using Grpahene, TiN and HOPG electrodes. Experimental details of the $MoS_2$ deposits including AFM, Raman and XPS measuremetns.

**TOC Figure**

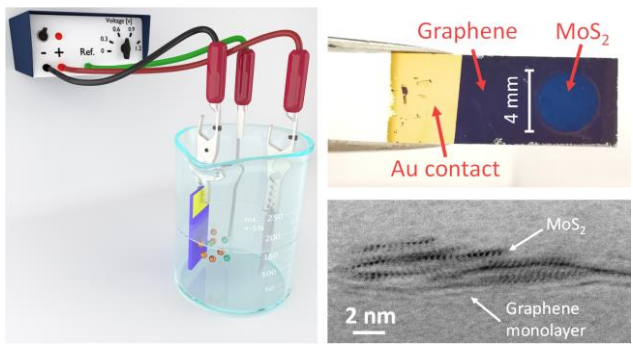